March 9, 1999
	Editors
	Physical Review D
	
	Dear Editors,
	              Please find below the TeX file of the
	paper titled "Effects of Kaluza-Klein Excitations 
	on $g_{\mu}-2$", by P. Nath and M. Yamaguchi which
	we are submitting for publication in Physical Review D.
	The post script file of two figures will be sent in a
	separate e-mail subsequent to this e-mail.
	
	Sincerely,
	Pran Nath              
        Department of Physics 
        Northeastern University 
        Boston, MA, 02115-5000

\documentstyle[osa,manuscript,graphicx]{revtex}

\begin{document}

\begin{center}{\Large \bf  Effects of Kaluza-Klein Excitations
on $g_{\mu}-2$ } \\
\vskip.25in
{Pran Nath$^a$ and Masahiro Yamaguchi$^b$ }

{\it
a Department of Physics, Northeastern University, Boston, MA 02115, USA\\
b Department of Physics, Tohoku University, Sendai 980-8578, Japan
} \\%

\end{center}

\begin{abstract}                
An analysis of the effects associated with the Kaluza-Klein 
excitations of the photon and of the W and Z bosons on
$g_{\mu}-2$ for d number of extra dimensions with
large radius compactifications is given. 
The Kaluza-Klein  effects on
 $g_{\mu}-2$ are found to be very 
 sensitive to the number of extra dimensions. For models where 
 the quark-leptons generations live on the 4D wall, it is shown
 that when the constraints from the Kaluza-Klein corrections   
  to the Fermi constant are included,  the 
 effects of the Kaluza-Klein excitations to $g_{\mu}-2$ become
 too small to be observable.
 A model which evades Kaluza-Klein corrections to the muon decay
 $\mu\rightarrow e\bar\nu_e\nu_{\mu}$  
 without suppression of their effects on $g_{\mu}-2$ is discussed.

\end{abstract}

\section{ Introduction}
The anomalous magnetic moment of the muon ($g_{\mu}$-2)
 is one of the most accurately determined 
quantities\cite{bailey} in
particle physics. Here we analyse the effects of 
Kaluza-Klein\cite{kaluza} excitations on $g_{\mu}$ in theories 
with large radius compactifications\cite{witten,anto,gf}.
Such theories are currently being investigated because they might
arise in the strong coupling limit of  the heterotic string 
which can yield a string scale in the TeV 
region\cite{witten}. The framework
of the model we work in is discussed in ref.\cite{gf}. 
Our view point is that
the fields of the minimal supersymmetric standard model reside
in (4+d) dimensions while gravity propagates in the 10 dimensional
bulk. In the language of Type I string the MSSM fields reside on
p=3+d branes and we consider  compactifications internal
to the brane. 

Our analysis is in the framework of an effective
field theory and we illustrate our procedure for the Standard 
Model (SM) case for d=5 .
 Here our SM fields will consist of gauge fields $A_M$ (M=0,1,2,3,5),
 the  Higgs multiplet $H$, and the
 quarks and lepton multiplets. 
 We assume a form of V(H) such that H develops a vacuum expectation value
and spontaneous breaking of the electro-weak symmetry takes place 
so that $H=\frac{1}{\sqrt 2}(V+H_1+iH_2)$.
We next choose  a five dimensional gauge fixing term of the form 
\begin{equation}
L_5^{gf}=-\frac{1}{2\xi}(\partial_MA^M-\xi Vg_5H_2)^2
\end{equation}
In this gauge the bilinear term involving $A_M$ and $H_2$ form a total
divergence of the  $\partial_M(H_2A^M)$ and hence can be dropped. 
The remaining terms in the Lagrangian have a simple decomposition among 
the vector bosons, the Higgs  and the fictitious Goldstone bosons.
 After spontaneous breaking we compactify the Lagrangian 
on $S^1/Z_2$ with a radius of 
compactification R.

 We shall assume that 
the Higgs and the gauge fields live in the five dimensional bulk and
the fermions live at one of the orbifold points.
Thus we will have only zero modes for the fermion fields in this case.
For the 4D Lagrangian we shall work in the $\xi=1$ gauge. In this
gauge, terms bilinear in $A_{\mu} (\mu=0,1,2,3)$
 and $A_5$ form a total divergence in 4 dimensions and
the $A_{\mu}$ and $A_5$ decouple at the  bilinear level. To normalize the 
4D Lagrangian a redefinition of the couplings is needed, i.e.,  
$g_5/{\sqrt{\pi R}}=g$.  Normal mode 
decomposition on  $S^1/Z_2$ indicates  mass terms for the W bosons
 of  $m_W^2+n^2/R^2$,  n=0,1,2,..,$\infty$,
where the first term arises from spontaneous breaking of the
electro-weak symmetry and the second term arises from the 
compactification, 
and similar relations hold for the Kaluza-Klein excitations of the
Z boson and for the Higgs boson. 
The above analysis
can be extended to include supersymmetry. Before compactification
this 5D theory has  an N=2 supersymmetry. After compactification
the massless sector of the theory has N=1 supersymmetry while the
massive Kaluza-Klein states maintain an N=2 supersymmetry.
The  model can be extended to d dimensions.

\section{Contribution of  W and Z Kaluza-Klein Excitations }
The framework of the analysis is as in ref.\cite{gf}.
The analysis of the one loop contribution
to $g_{\mu}-2$ can be carried out in the usual fashion and
a finite contribution arises from each Kaluza-Klein mode. 
Neglecting the relatively small contributions from the Yukawa 
couplings, the contributions to $g_{\mu}-2$ at the one loop level
from the W and Z exchange and from their Kaluza-Klein excitations 
for d extra dimensions is given by 

\begin{equation}
(\Delta a)_{\mu}= \frac{G_F^{SM} m_{\mu}^2}{\pi^22\sqrt 2}
(\frac{5}{6} K_d(\frac{M_W}{M_R})
+(-\frac{5}{12}+\frac{4}{3}(sin^2\theta_W-\frac{1}{4})^2) K_d(\frac{M_Z}{M_R}))
\end{equation}
Here $a_{\mu}$=($g_{\mu}$-2)/2, 
 $G_F^{SM}$ is  the Fermi constant as evaluated in the 
Standard Model, $\theta_W$   is the weak angle,
 $M_W$ ($M_Z$) is the W (Z) boson mass, $M_R=1/R$ is the
 common mass scale at 
 which the extra dimensions open up,  and 
 the function $K_d(c)$ is defined by
\begin{equation}
K_d(c)= \int_{0}^{\infty}dt e^{-t}  
(\theta_3(\frac{it}{c\pi}))^{d}
\end{equation}
Here $\theta_3(\tau)$ for complex $\tau$ is given by $\theta_3(\tau)$=
$\sum_{k=-\infty}^{\infty}$$exp(i\pi k^2 \tau)$ where  ($Im\tau>0$) . 
In the limit c=0 one has $K_d(0)=1$ and  Eq.(2) reduces to the Standard 
Model result\cite{fuji}. For the case d=1 Eq.(3) gives a convergent 
result. However, a cuttoff $\Lambda_d$ on the lower limit of the
integral in Eq.(3) is needed  for the case $d\geq 2$. As discussed
in Ref.\cite{gf} this cutoff is determined consistently by comparison
with the truncation on the sum over the Kaluza-Klein states so that
only the states up to the string scale are retained in the sum.
One finds the following results for $\Lambda_d$ for $d\geq 2$.
\begin{eqnarray}
\Lambda_2=(\frac{M_R}{M_{str}})^2\nonumber\\
\Lambda_d=
(\Gamma(\frac{d}{2}))^{\frac{2}{d-2}}(\frac{M_W}{M_{str}})^2,~~d\geq 3
\end{eqnarray}

As discussed in Ref.\cite{gf} one must carry out a
 redefinition of the Fermi constant 
because of the dressing of the Kaluza-Klein states and the 
experimentally observed $G_F$ is to be identified as 

\begin{equation}
G_F=G_F^{SM}K_d(\frac{M_W^2}{M_R^2})
\end{equation}
Using this redefinition  we find 
 the contribution of the Kaluza-Klein modes to $a_{\mu}$ is 
\begin{equation}
(\Delta a)_{\mu}^{W-ZKK}= \frac{G_F m_{\mu}^2}{\pi^22\sqrt 2}
(-\frac{5}{12}+\frac{4}{3}(sin^2\theta_W-\frac{1}{4})^2) 
(K_d(\frac{M_Z}{M_R})- K_d(\frac{M_W}{M_R})  )/ K_d(\frac{M_W}{M_R}))
\end{equation}
For d=1  $K_1(c)$ is given approximately by 
$K_1(c)\simeq 1+\frac{\pi^2}{3} c$ while the $d\geq 2$ cases require a cutoff
as discussed above and one finds that the following results 
give a good approximation

\begin{eqnarray}
K_2(c)=1+c(\frac{2\pi^2}{3}+2\pi ln\frac{M_{str}}{M_R})\nonumber\\ 
K_d(c)(d\geq 3)= 1+
 c(\frac{d}{d-2})\frac{\pi^{d/2}}{\Gamma(1+\frac{d}{2})}
(\frac{M_{str}}{M_R})^{d-2} 
\end{eqnarray}

\section{Contribution of Photonic Kaluza-Klein Excitations}
In the above we computed the contributions arising from the W and Z  
Kaluza-Klein excitations to $a_{\mu}$. There are also 
contributions to  $a_{\mu}$ arising from the exchange of 
the Kaluza-Klein excitations of the photon. Here our analysis 
gives for the d extra dimensions the result

\begin{equation}
(\Delta a)_{\mu}^{\gamma KK}= \frac{\alpha}{3\pi} 
\frac{m_{\mu}^2}{M_R^2} G
\end{equation}
where 
\begin{equation}
G=\int_{0}^{\infty} dt (\theta_3(\frac{it}{\pi}))^d-1
\end{equation} 
As for the case of the W and Z Kaluza-Klein exchange contributions,
 for d=1 the photonic Kaluza-Klein exchange contribution 
  is finite  while a cutoff is needed for
higher values of d. The cutoff is introduced in
the same way as discussed for the case $K_d$ and one gets the following 
results for various cases 

\begin{equation}
(\Delta a)_{\mu}^{\gamma KK} (d=1)= \alpha \frac{\pi}{9} 
\frac{m_{\mu}^2}{M_R^2}
\end{equation}
\begin{equation}
(\Delta a)_{\mu}^{\gamma KK} (d=2)= \frac{\alpha}{3\pi}
(\frac{2\pi^2}{3}+2\pi ln\frac{M_{str}}{M_R}) 
\frac{m_{\mu}^2}{M_R^2}
\end{equation}
\begin{equation}
(\Delta a)_{\mu}^{\gamma KK} (d\geq 3)= \frac{\alpha}{3\pi} 
(\frac{d}{d-2})\frac{\pi^{d/2}}{\Gamma(1+\frac{d}{2})}
(\frac{M_{str}}{M_R})^{d-2}
\frac{m_{\mu}^2}{M_R^2} 
\end{equation}
The total Kaluza-Klein exchange contribution to $a_{\mu}$ is the
sum of the contribution from the  W and Z Kaluza-Klein 
exchange contribution and from the photonic W and Z exchange 
contribution
\begin{equation}
a_{\mu}^{KK}=a_{\mu}^{W-ZKK}+a_{\mu}^{\gamma KK}
\end{equation}
We note that after redefinition of the Fermi constant 
the W and Z Kaluza-Klein exchange contribution $a_{\mu}^{W-ZKK}$
as given by Eq.(6) is negative while the photonic Kaluza-Klein 
exchange contribution
 $a_{\mu}^{\gamma KK}$ as given by Eq.(8) is positive. Thus one has a partial 
 cancellation between these two contributions. We emphasize here that in
 any numerical analysis of Eq.(13) the constraints arising 
 from Eq.(5) (which holds only within the current error bars 
 of $G_F^{SM}$ and experiment on $G_F$) should be included. 
 For the compactifications considered  here and
 in Ref.\cite{gf} these constraints on 
$M_R$ at the $2\sigma$ level are $M_R>1.6$ TeV\cite{gf,mp}
	for d=1, $M_R>3.5$ TeV for d=2, $M_R>5.7$ TeV for d=3 and
	$M_R>7.8$ TeV for d=4.

\section{ Sizes of Kaluza-Klein Effects on $(g_{\mu}-2)$}  
	We discuss now the prospects for the observation of the 
	Kaluza-Klein contributions in $g_{\mu}$ experiment. 
	For this purpose we review 
	first the current situation on $a_{\mu}$. 
	The current experimental value of $a_{\mu}$ is ~\cite{bailey}:
$a_{\mu}^{exp}=11659230(84)\times 10^{-10}$, where the 
quantity in the parenthesis is the uncertainty while
the corresponding uncertainty in the theoretical determinations is
 significantly 
 smaller. The most recent Standard model prediction for $a_{\mu}$
is $a_{\mu}^{theory}(SM)=11659162.8(6.5)\times 10^{-10}$.
This result includes 
 $\alpha^5$ QED contributions\cite{kino}, 
 the hadronic vacuum polarization\cite{davier}
 and  light by light hadronic contributions\cite{hayakawa}, 
 and the Standard Model 
 electro-weak contributions\cite{czar}. 
 Essentially all of the uncertainty in the Standard Model result 
 arises from the hadronic contributions which 
 include the hadronic vacuum 
polarization~\cite{davier}, and 
the light by light hadronic contribution~\cite{hayakawa}.
The total Standard Model electro-weak contribution 
 up to two loops~\cite{czar} is given by  
$a_{\mu}^{EW}(SM)=15.1(0.4)\times 10^{-10}$. 

The new $g_{\mu}$
 Brookhaven experiment E821\cite{hertzog}
will be able to reduce the current experimental uncertainty 
by a factor of
20 to a level of $4\times 10^{-10}$.
With this sensitivity the new $g_{\mu}$ experiment will be able to test the
Standard Model electro-weak contribution even with the current level of
uncertainty in the hadronic contributions. 
Infact the experiment will be able to probe $a_{\mu}$ to a level of
$\sim 8\times 10^{-10}$ (where we have combined the experimental error
and the 
hadronic error in quadrature). 
 Further, it is 
expected that the uncertainty in the hadronic error may reduce 
even more by perhaps as much as a factor of 2 from the data from the
ongoing and future precision low energy experiments at  
VEPP-2M, 
DA$\Phi$NE and BEPC. Of course a 
further reduction of the hadronic error will sharpen the 
ability of the Brookhaven experiment to probe new physics.

	 In the analysis  of $a_{\mu}^{KK}$ we  
	impose the gauge coupling unification to constrain the ratio 
	$M_{str}/M_R$ the details of which are discussed in ref.\cite{gf}.
	The results are displayed in Fig.1. The analysis exhibits 
	a sharp dependence of the contribution of the Kaluza-Klein
	excitations to $a_{\mu}^{KK}$ on the number of extra space time 
	dimensions. The sharp dependence on d arises from the 
	summation over the number of Kaluza-Klein states because
	of the combinatorics factors. As discussed already there is
	a partial cancellation between the W and Z Kaluza-Klein
	exchange contribution and the photonic Kaluza-Klein
	exchange contribution after a redefinition of the Fermi
	constant is taken into account. 
	As discussed above and in Ref.\cite{gf} the analysis 
	of the Kaluza-Klein
	mode contribution to $G_F$ gives strong constraints on
	$M_R$. Fig.1 shows that with these constraints 
	on $M_R$ the new $g_{\mu}$ experiment will not come
	even close to exploring the extra dimensions, ie., the 
	total Kaluza-Klein contribution falls  more than 
	 1-2 orders of magnitude below the sensitivity 
	that will be achievable in the
	new $g_{\mu}$ experiment.
	
		In the above we assumed that all the quark-lepton
		generations lie on the 4-dimensional wall. We 
		discuss now a variant of the model considered 
		above  where the 
		first quark-lepton generation lies on the wall
		while the second quark-lepton generation lives
		in the  bulk. In this case the first quark-lepton
		generation will have no couplings with the 
		Kaluza-Klein modes while the second generation 
		will have such couplings. In this model the 
		process $\mu\rightarrow e\bar\nu_e\nu_{\mu}$ receives
		no Kaluza-Klein correction at the tree level and
		thus $G_F$ is uncorrected at this level. Similarly
		in this case the atomic parity violating interactions
		receive no Kaluza-Klein corrections. 
		Consequently there
		are no  constraints arising at the tree level 
		from the experimental accuracy
		of $G_F$ and from the atomic parity experiments on 
		$M_R$, although corrections at the loop level can arise 
		which are, however, expected to be much smaller
		than what one would otherwise expect. 
		Other experimental bounds as those arising from contact
		interactions using LEP data are also evaded in this 
		model. Further, one would also expect here a significant 
		suppression of the production of the Kaluza-Klein excitations
		of the W and the Z boson at the $\bar p p$
		colliders due to the 
		dominance of the first generation in the quark content
		of p($\bar p$). 
		However,  Kaluza-Klein corrections to
		$g_{\mu}-2$ in this model are  not suppressed  
		and the analysis of
		$a_{\mu}^{KK}$ in this case is given in Fig.2 
		again 
		under the unification of the gauge coupling constant
		constraint. (The analysis uses Eq.2 with the measured
		value of $G_F^{SM}$). The analysis of Fig.2 shows that   
		the new $g_{\mu}$ experiment will
		be able to probe two extra dimensions
		up to scales $M_R\sim 0.65$ TeV, three extra dimensions
		up to scales $M_R\sim 1$ TeV, and four extra dimensions
	        up to scales $M_R\sim 1.4$ TeV. We note that the scales
	        which can be explored well exceed the mass bound 
	  	 on the
	        Kaluza-Klein modes of the quarks inferred from the
	        fourth generation quark searches which gives
	        a lower limit of $130$ GeV\cite{fsn}.

\section{Conclusion}
In this paper we have investigated the effects of Kaluza-Klein 
excitations arising from extra dimensions with large radius 
compactifications  on
$g_{\mu}-2$. These effects consist of the photonic Kaluza-Klein 
exchange contribution and of the  W and  Z exchange contribution.
For the case of one extra dimension the Kaluza-Klein 
contribution to $a_{\mu}$ is finite. However, a cutoff is 
necessary for the case of d extra dimensions, $d\geq 2$. 
We have derived approximate relations for the Kaluza-Klein  
contributions for these cases (see Eqs. 6,7,10-12). 
For d=2 these approximate relations are accurate to $O(10\%)$ while for
$d\geq 3$ the relations are accurate to $O(1-2\%)$. We have 
given a quantitative analysis of the contributions of 
the Kaluza-Klein excitations of $\gamma$, W and Z under the
constraints of the unification of the gauge coupling 
constants.  Our analysis shows that for the case 
when all the quark-lepton
generations  lie on the 4D wall, inclusion of the constraints 
on $M_R$ arising from the analysis of Kaluza-Klein mode contributions
to $G_F$  lead to effects on $g_{\mu}-2$ which are too small to
be observable by the new $g_{\mu}$ experiment. We also considered
a  model where the first generation lives in the bulk and 
the second generation lies on the wall. In this case effects 
from extra dimensions $d\geq 2$ may become 
 visible in the Brookhaven experiment. Furthermore, a muon collider
 may be able to explore these extra dimensions directly. 
Of course a fundamental string model will have to justify such an
assignment of generations between the bulk and the wall.

 After completion of this work there appeared a paper\cite{graesser}
 where an analysis of the effects of 
 Kaluza-Klein excitations of gauge bosons on $g_{\mu}-2$ is also given. 
 However, the constraints arising from the effects of Kaluza-Klein
 corrections to the Fermi constant are not included. 

\acknowledgments
The research of PN was supported in part by the National Science Foundation
grant no. PHY-9602074.
The work of MY was supported in part by the Grant-in-Aid for
Scientific Research from the Ministry of Education of Japan on
Priority Area 707 "Supersymmetry and Unified Theory of Elementary
Particles", and by the Grant-in-Aid No.09640333.

\begin{figure}
\begin{center}
\includegraphics[angle=0,width=4.5in]{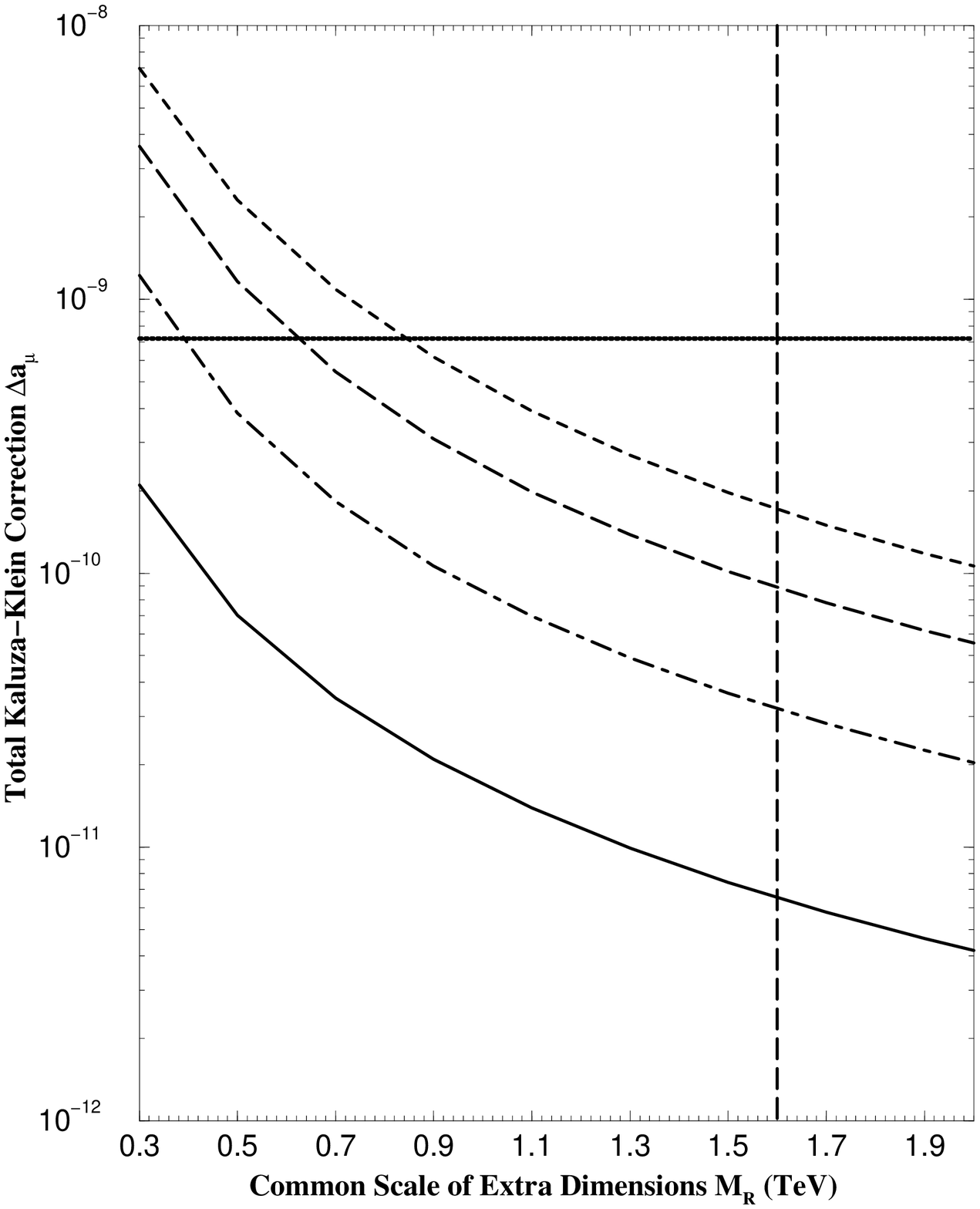}
\caption{Plot of sum of the photonic Kaluza-Klein exchange contribution 
and the W and Z Kaluza-Klein exchange contribution to 
 $a_{\mu}$ as a function
of the common scale of extra dimension when the $G_F$ constraint is
included. 
The curves correspond to the
case d=1 (solid), d=2 (dot-dashed), d=3 (long dashed), and d=4
(dashed). The horizontal dotted line corresponds to the 1$\sigma$ 
limit to which the Brookhaven experiment E821 will be able to probe
$a_{\mu}$. The $G_F$ constraint requires $M_R>1.6$ TeV for d=1 and the
allowed region lies to the right of the vertical dashed curve.
The constraints on $M_R$ for $d>1$ are even more severe.} 
\label{fig1}
\end{center}
\end{figure}

\begin{figure}
\begin{center}
\includegraphics[angle=0,width=4.5in]{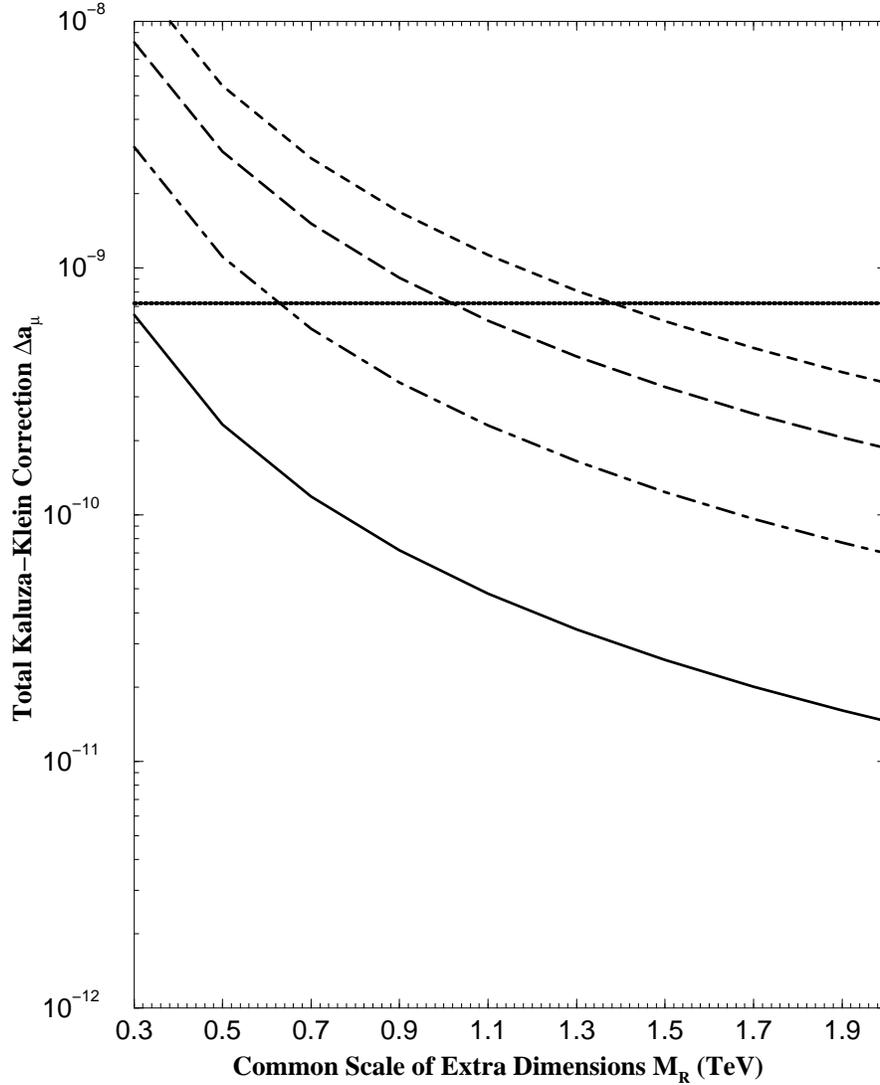}
\caption{Plot of sum of photonic Kaluza-Klein exchange contribution 
and the W and Z Kaluza-Klein exchange contribution to 
 $a_{\mu}$ as a function
of the common scale of extra dimension when there are no Kaluza-Klein
corrections to $G_F$. 
The curves correspond to the
case d=1 (solid), d=2 (dot-dashed), d=3 (long dashed), and d=4
(dashed). The horizontal dotted line is the same as in Fig.1.}
\label{fig2}
\end{center}
\end{figure}

\end{document}